\documentclass[
superscriptaddress,floatfix]{revtex4}
\usepackage{graphics}
\usepackage{latexsym,amssymb,amsmath}
\usepackage{amscd,array,dsfont,graphicx,kbordermatrix,mathtools,multirow,rotating,subfig,tensor,times,verbatim,youngtab}

\usepackage{epsfig}
\usepackage{amsmath}
\usepackage{amssymb}
\usepackage{calc}
\usepackage[bookmarksnumbered,bookmarksopen=true,bookmarksopenlevel=1,linktocpage,pdfstartview=FitH]{hyperref}
\usepackage[all]{hypcap}

\newcolumntype{M}[1]{>{$}{#1}<{$}}

\newcommand{\beq}{\begin{equation}}
\newcommand{\eeq}{\end{equation}}
\newcommand{\bea}{\begin{eqnarray}}
\newcommand{\be}{\begin{equation}}
\newcommand{\ee}{\end{equation}}
\newcommand{\eea}{\end{eqnarray}}
\newcommand{\ba}{\begin{array}}
\newcommand{\ea}{\end{array}}
\newcommand{\bit}{\begin{itemize}}
\newcommand{\eit}{\end{itemize}}
\newcommand{\ben}{\begin{enumerate}}
\newcommand{\een}{\end{enumerate}}

\allowdisplaybreaks[1]

\newcommand{\rep}[1]{\ensuremath{\mathbf{#1}}}

\newcommand{\sfx}{\textsf{x}}
\newcommand{\sfo}{\textsf{o}}

\newcolumntype{D}[1]{>{$\displaystyle}{#1}<{$}}

\newcolumntype{C}[1]{>{\centering} m{#1}}
\newcolumntype{X}[1]{>{\centering $} m{#1}<{$}}

\begin{document}

\author{ M. J. Duff }
\affiliation{The Blackett Laboratory, Imperial College London, Prince Consort Road, London SW7 2BZ, U.K.}
\title{Black holes and qubits
\footnote{Talk delivered at the Pontifical Academy of Sciences Symposium on Subnuclear Physics, Vatican City, October 2011.}
}

\begin{abstract}
{Quantum entanglement lies at the heart of quantum information theory, with applications to quantum computing, teleportation, cryptography and communication. In the apparently separate world of quantum gravity, the Hawking effect of radiating black holes has also occupied centre stage. Despite their apparent differences, it turns out that there is a correspondence between the two.}
\end{abstract}
\maketitle{}

\renewcommand\thesection{\arabic{section}}
\renewcommand\thesubsection{\arabic{subsection}}

\numberwithin{subsection}{section}
\numberwithin{subsubsection}{subsection}

\section{Introduction}

Whenever two very different areas of theoretical physics are found to share the same mathematics, it frequently leads to new insights on both sides. Here we describe how knowledge of string theory and M-theory leads to new discoveries about Quantum Information Theory (QIT) and vice-versa \cite{Duff1,Kalloshlinde,Levay1}. 

\section{Bekenstein-Hawking entropy} 

Every object, such as a star, has a critical size determined by its mass, which is called the Schwarzschild radius. A black hole is any object smaller than this. Once something falls inside the Schwarzschild radius, it can never escape. This boundary in spacetime is called the event horizon. So the classical picture of a black hole is that of a compact object whose gravitational field is so strong that nothing, not even light, can escape. 

Yet in 1974 Stephen Hawking showed that quantum black holes are not entirely black but may radiate energy, due to quantum mechanical effects in curved spacetime. In that case, they must possess the thermodynamic quantity called entropy. Entropy is a measure of how organized or disorganized a system is, and, according to the second law of thermodynamics, it can never decrease. Noting that the area of a black hole event horizon can never decrease, Jacob Bekenstein had earlier suggested such a thermodynamic interpretation implying that black holes must have entropy. This Bekenstein-Hawking black hole entropy is in fact given by one quarter the area of the event horizon. This is a remarkable fact relating a thermodynamic quantity, entropy, with a quantum mechanical origin, to a purely geometrical quantity, area, that is calculated in Einstein's classical theory of gravity.

Entropy also has a statistical interpretation as a measure of the number of quantum states available. However, it was not until 20 years later that string theory, as a theory of quantum gravity, was able to provide a microscopic explanation of this kind.

\section{Bits and pieces} 

A classical bit is the basic unit of computer information and takes the value 0 or 1. A light switch provides a good analogy; it can either be off, denoted 0, or on, denoted 1. A quantum bit or ``qubit'' can also have two states but whereas a classical bit is either 0 or 1, a qubit can be both 0 and 1 until we make a measurement. In quantum mechanics, this is called a superposition of states. When we actually perform a measurement, we will find either 0 or 1 but we cannot predict with certainty what the outcome will be; the best we can do is to assign a probability. 

There are many different ways to realize a qubit physically. Elementary particles can carry an intrinsic spin. So one example of a qubit would be a superposition of an electron with spin up, denoted 0, and an electron with spin down, denoted 1. Another example of a qubit would be the superposition of the left and right polarizations of a photon. So a single qubit state, usually called Alice, is a superposition of Alice-spin-up 0 and Alice-spin-down 1, represented by the line in figure 1. The most general two-qubit state, Alice and Bob, is a superposition of Alice-spin-up-Bob-spin-up 00, Alice-spin-up-Bob-spin-down 01, Alice-spin-down-Bob-spin-up 10 and Alice-spin-down-Bob-spin-down 11, represented by the square in figure 1. 
\begin{figure}[pth!]
\centerline{\includegraphics[width=12.0truecm,clip=]{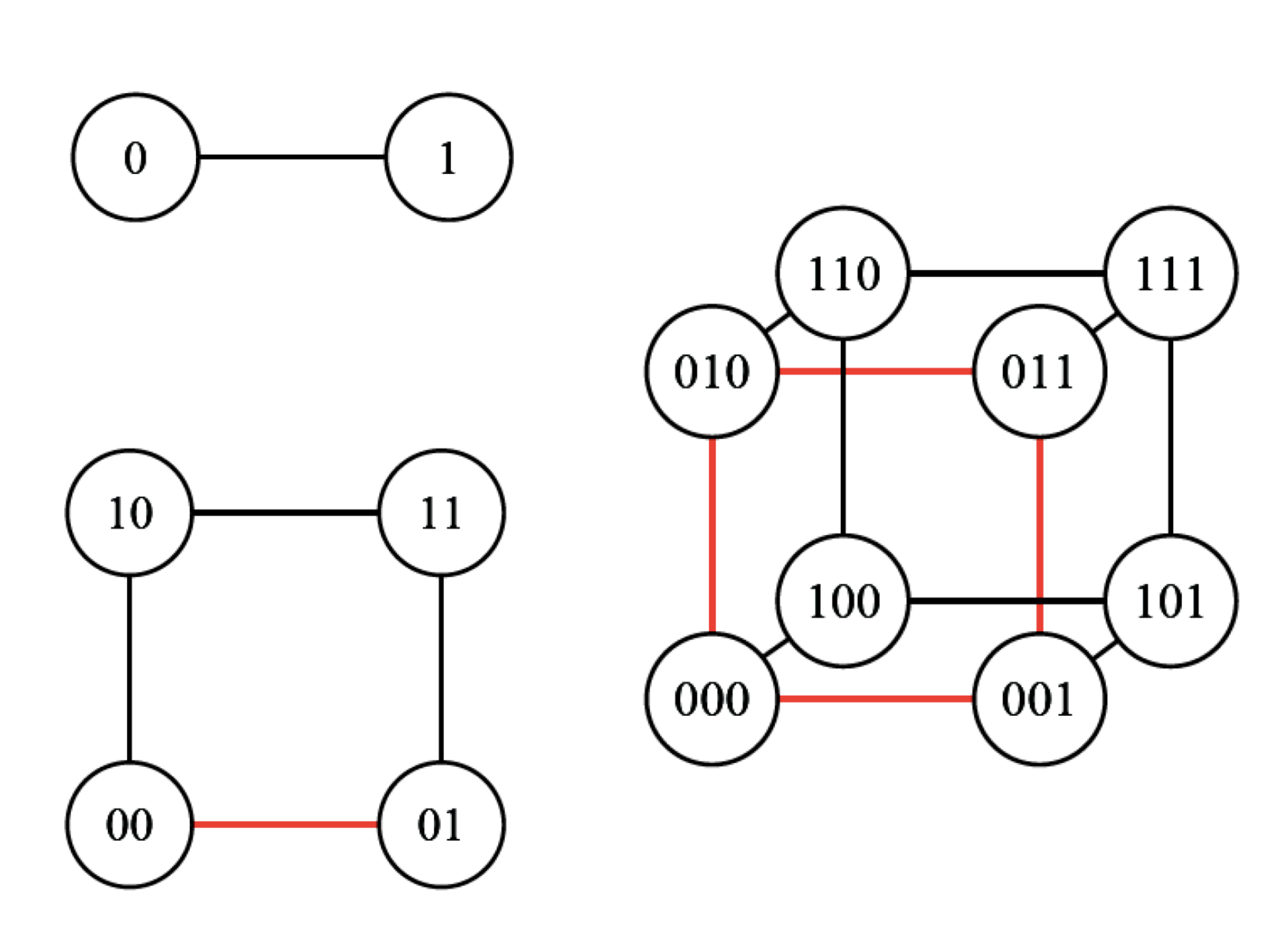}}
\caption{A single quit is represented by a line, two qubits by a square and three qubits by a cube.}
\end{figure}

Consider a special two-qubit state which is just $00 + 01$. Alice can only measure spin up but Bob can measure either spin up or spin down. This is called a separable state; Bob's measurement is uncorrelated with Alice's. By contrast consider $00 + 11$. If Alice measures spin up, so must Bob and if she measures spin down so must he. This is called an entangled state; Bob cannot help making the same measurement. Mathematically, the square in figure 1 forms a $2 \times 2$ matrix and a state is entangled if the matrix has a nonzero determinant. 

This is the origin of the famous Einstein-Podolsky-Rosen (EPR) paradox put forward in 1935. Even if Alice is in the Vatican and Bob is millions of miles away in Alpha Centauri, Bob's measurement will still be determined by Alice's. No wonder Albert Einstein called it ``spooky action at a distance''. EPR concluded rightly that if quantum mechanics is correct then nature is nonlocal and if we insist on local ``realism'' then quantum mechanics must be incomplete. Einstein himself favoured the latter hypothesis. However, it was not until 1964 that CERN theorist John Bell proposed an experiment that could decide which version was correct, and it was not until 1982 that Alain Aspect actually performed the experiment. Quantum mechanics was right, Einstein was wrong and local realism went out the window. 

As QIT developed, the impact of entanglement went far beyond the testing of the conceptual foundations of quantum mechanics. Entanglement is now essential to numerous quantum information tasks such as quantum cryptography, teleportation and quantum computation. 

\section{Cayley's hyperdeterminant} 

As a high-energy theorist involved in research on quantum gravity, string theory and M-theory, I paid little attention to all this, even though as a member of staff at CERN in the 1980s my office was just down the hall from Bell's. 

My interest was not aroused until 2006, when I attended a lecture by Hungarian physicist Peter Levay at a conference in Tasmania. He was talking about three qubits Alice, Bob and Charlie where we have eight possibilities 000, 001, 010, 011, 100, 101, 110, 111, represented by the cube in figure 1. Wolfgang Dur and colleagues at the University of Innsbruck have shown \cite{Dur} that three qubits can be entangled in several physically distinct ways: tripartite GHZ (Greenberger-Horne-Zeilinger), tripartite W, biseparable A-BC, separable A-B-C and null, as shown in the right hand diagram of figure 2. 
\begin{figure}[pth!]
\centerline{\includegraphics[width=\textwidth]{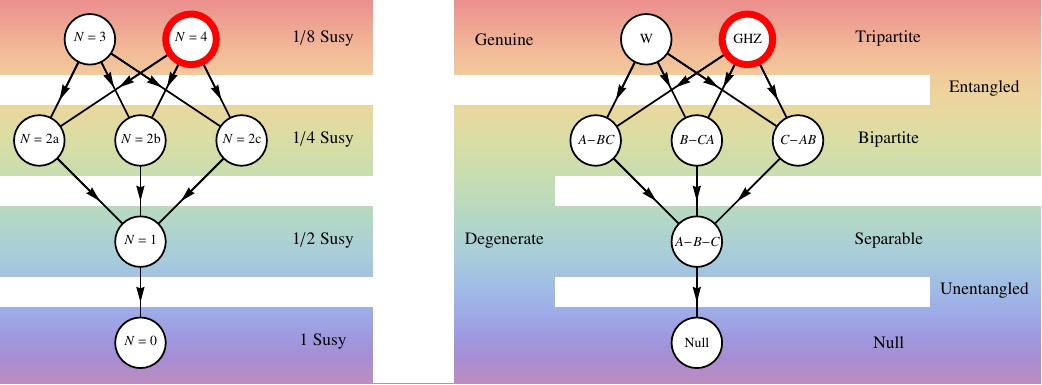}}
\caption{The classification of black holes from $N$ wrapped branes (left) exactly matches the classification of three-qubit entanglement (right). Only the GHZ state has non-zero 3-tangle and only the $N=4$ black hole has non-zero entropy.}
\end{figure}

The GHZ state is distinguished by a nonzero quantity known as the 3-tangle, which measures genuine tripartite entanglement. Mathematically, the cube in figure 1 forms what in 1845 the mathematician Arthur Cayley called a $2 \times 2 \times 2$ hypermatrix and the 3-tangle is given by the generalization of a determinant called Cayley's hyperdeterminant. 

The reason this sparked my interest was that Levay's equations reminded me of some work I had been doing on a completely different topic in the mid 1990s with my collaborators Joachim Rahmfeld and Jim Liu \cite{Duffstu}. We found a particular black hole solution that carries eight charges (four electric and four magnetic) and involves three fields called $S, T$ and $U$. When I got back to London from Tasmania I checked my old notes and asked what would happen if I identified $S, T$ and $U$ with Alice, Bob and Charlie so that the eight black-hole charges were identified with the eight numbers that fix the three-qubit state. I was pleasantly surprised to find that the Bekenstein-Hawking entropy of the black holes was given by the 3-tangle: both were described by Cayley's hyperdeterminant.
This turned out to be the tip of an iceberg and there is now a growing dictionary between phenomena in the theory of black holes and phenomena in QIT.

\section{Octonions} 

According to supersymmetry, for each known boson (integer spin 0, 1, 2 and so on), there is a fermion (half-integer spin 1/2, 3 /2, 5/2 and so on), and vice versa. CERN's Large Hadron Collider will be looking for these superparticles. The number of supersymmetries is denoted by $\mathcal{N}$ and ranges from 1 to 8 in four spacetime dimensions. 

CERN's Sergio Ferrara and I have extended the $STU$ model example, which has $\mathcal{N} =2$, to the most general case of black holes in $\mathcal{N} =8$ supergravity. We have shown that the corresponding system in quantum information theory is that of seven qubits (Alice, Bob, Charlie, Daisy, Emma, Fred and George), undergoing at most a tripartite entanglement of a very specific kind as depicted by the Fano plane of figure 3. The Fano plane has a strange mathematical property: it describes the multiplication table of a particular kind of number: the octonion. Mathematicians classify numbers into four types: real numbers, complex numbers (with one imaginary part A), quaternions (with three imaginary parts $A, B, D$) and octonions (with seven imaginary parts $A, B, C, D, E, F, G$). Quaternions are non-commutative because $AB$ does not equal $BA$. Octonions are not only noncommutative but also non-associative since $(AB)C$ does not equal $A(BC)$. 
\begin{figure}[pth!]
\centerline{\includegraphics[width=10.0truecm,clip=]{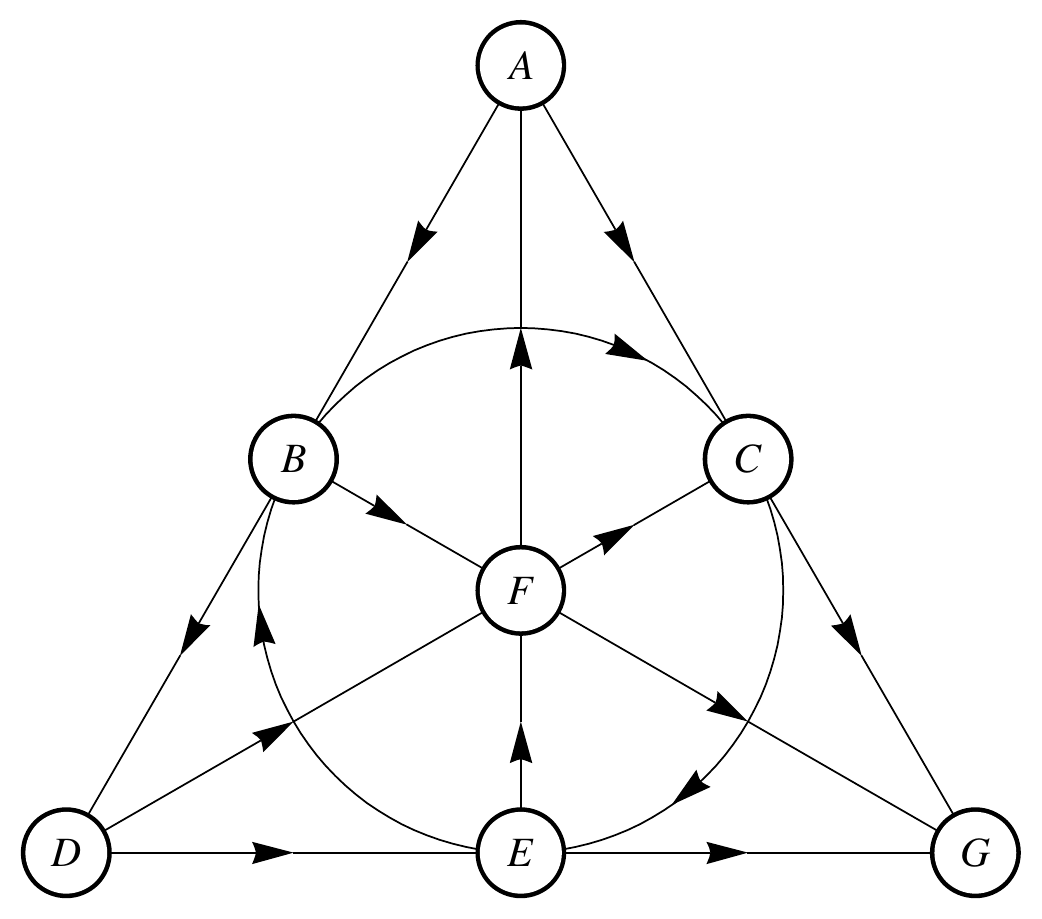}}
\caption{The Fano plane. The vertices $A,B,C,D,E,F,G$ represent the seven qubits and the seven lines $ABD$, $BCE$, $CDF$, $DEG$, $EFA$, $FGB$, $GAC$, represent the tripartite entanglement.}
\end{figure}

Real, complex and quaternion numbers show up in many physical contexts. Quantum mechanics, for example, is based on complex numbers and Pauli's electron spin operators are quaternionic. Octonions have fascinated mathematicians and physicists for decades but have yet to find any physical application. In recent books both Roger Penrose and Ray Streater have characterized octonions as one of the great “lost causes” in physics. So we hope that the tripartite entanglement of seven qubits (which is just at the limit of what can be reached experimentally) will prove them wrong and provide a way of seeing the effects of octonions in the laboratory \cite{Duff2,Levayfano,Borstenrev}. 

\section{Implications for M-theory}

We have also learned things about M-theory from QIT. The Fano plane suggests a whole new way of studying its symmetries based on the 7 imaginary octonions (completely different from the
Jordan algebra approach that uses all 8 split octonions). Such expectations have recently been strengthened by
the discovery of four supergravities with $16+16, 32+32, 64+64, 128+128$ degrees of freedom displaying some curious properties \cite{Duffferrara1} . In particular they reduce to $\mathcal{N} = 1; 2; 4; 8$ theories all with maximum rank 7 in $D=4$ which correspond to 0, 1, 3, 7 lines of the Fano plane and hence admit a division algebra $(\mathds{R;C;H;O})$ interpretation consistent with the black-hole/qubit correspondence. They exhibit unusual properties. For example they are all self-mirror with vanishing trace anomaly \cite{Duffferrara2} .  

\section{Superqubits}

In another development, QIT has been extended to super-QIT with the introduction of the superqubit which can take on three values: 0 or 1 or \$. Here 0 and 1 are “bosonic” and \$ is “fermionic” \cite{Borstensuper}. Such values can be realised in condensed matter physics, such as the excitations of the t-J model of strongly correlated electrons, known as spinons and holons. The superqubits promise totally new effects, for example, could they be even more non-local than ordinary bits? Super quantum computing is also being investigated \cite{Castellani}.

\section{Wrapped branes as qubits}

If current ideas are correct, a unified theory of all physical phenomena will require some radical ingredients in addition to supersymmetry. For example, there should be extra dimensions: supersymmetry places an upper limit of 11 on the dimension of spacetime. The kind of real, four-dimensional world that supergravity ultimately predicts depends on how the extra seven dimensions are rolled up, in a way suggested by Theodor Kaluza and Oskar Klein in the 1920s. In 1984, however, 11-dimensional supergravity was knocked off its pedestal by superstring theory in 10 dimensions. There were five competing theories: the E8 x E8 heterotic, the SO(32) heterotic, the SO(32) Type I, and the Type IIA and Type IIB strings. The E8 x E8 seemed, at least in principle, capable of explaining the elementary particles and forces, including their handedness. Moreover, strings seemed to provide a theory of gravity consistent with quantum effects. 

However, the spacetime of 11 dimensions allows for a membrane, which may take the form of a bubble or a two-dimensional sheet. In 1987 Howe, Inami, Stelle and I were able to show \cite{Howe} that if one of the 11 dimensions were a circle, we could wrap the sheet around it once, pasting the edges together to form a tube. If the radius becomes sufficiently small, the rolled-up membrane ends up looking like a string in 10 dimensions; it yields precisely the Type IIA superstring. In a landmark talk at the University of Southern California in 1995, Edward Witten \cite{Witten} drew together all of this work on strings, branes and 11 dimensions under the umbrella of M-theory in 11 dimensions. Branes now occupy centre stage as the microscopic constituents of M-theory, as the higher-dimensional progenitors of black holes and as entire universes in their own right. 

Such breakthroughs have led to a new interpretation of black holes as intersecting black-branes wrapped around the seven curled dimensions of M-theory or six of string theory. Moreover, the microscopic origin of the Bekenstein-Hawking entropy is now demystified. Using Polchinski's D-branes, Andrew Strominger and Cumrun Vafa were able to count the number of quantum states of these wrapped branes \cite{Strominger}. A p-dimensional D-brane (or Dp-brane) wrapped around some number p of the compact directions$(x_4, x_5, x_6, x_7, x_8, x_9)$ looks like a black hole (or D0-brane) from the four-dimensional $(x_0, x_1, x_2, x_3)$ perspective. Strominger and Vafa found an entropy that agrees with Hawking's prediction, placing another feather in the cap of M-theory. Yet despite all these successes, physicists are glimpsing only small corners of M-theory; the big picture is still lacking. Over the next few years we hope to discover what M-theory really is. Understanding black holes will be an essential pre-requisite. 

In string literature one may find D-brane intersection rules that tell us how N branes can intersect over one another and the fraction of supersymmetry that they preserve. In our black hole/qubit correspondence, my students Leron Borsten, Duminda Dahanayake, Hajar Ebrahim, William Rubens and I showed that the microscopic description of the GHZ state, $000 +011+101+110$ is that of the $N = 4$, fraction 1/8, case of D3-branes of Type IIB string theory \cite{wrapornottowrap}. We denoted the wrapped circles by crosses and the unwrapped circles by noughts; $0$ corresponds to $\sfx\sfo$ and 1 to $\sfo\sfx$, as in table 1. So the number of qubits here is three because the number of extra dimensions is six. This also explains where the two-valuedness enters on the black-hole side. To wrap or not to wrap; that is the qubit.

\begin{table}[ht]
\begin{tabular*}{\textwidth}{@{\extracolsep{\fill}}*{11}{c}>{$\lvert}c<{\rangle$}c}
\toprule
& 4    & 5    & & 6    & 7    & & 8    & 9    & macro charges & micro charges              & ABC & \\
\colrule
& \sfx & \sfo & & \sfx & \sfo & & \sfx & \sfo & $p^0$         & 0                          & 000 & \\
& \sfo & \sfx & & \sfo & \sfx & & \sfx & \sfo & $q_1$         & 0                          & 110 & \\
& \sfo & \sfx & & \sfx & \sfo & & \sfo & \sfx & $q_2$         & $-N_3\sin\theta\cos\theta$ & 101 & \\
& \sfx & \sfo & & \sfo & \sfx & & \sfo & \sfx & $q_3$         & $N_3\sin\theta\cos\theta$  & 011 & \\
\colrule
& \sfo & \sfx & & \sfo & \sfx & & \sfo & \sfx & $q_0$         & $N_0+N_3\sin^2\theta$      & 111 & \\
& \sfx & \sfo & & \sfx & \sfo & & \sfo & \sfx & $-p^1$        & $ -N_3\cos^2\theta$        & 001 & \\
& \sfx & \sfo & & \sfo & \sfx & & \sfx & \sfo & $-p^2$        & $-N_2$                     & 010 & \\
& \sfo & \sfx & & \sfx & \sfo & & \sfx & \sfo & $-p^3$        & $-N_1$                     & 100 & \\
\botrule
\end{tabular*}
\caption[Wrapped D3-branes]{Three qubit interpretation of the 8-charge $D=4$ black hole from four D3-branes wrapping around the lower four cycles of $T^6$ with wrapping numbers $N_0,N_1,N_2,N_3$ and then allowing $N_3$ to intersect at an angle $\theta$.}\label{tab:3QubitIntersect}
\end{table}

\section{Repurposing string theory}

  In the forty years since its inception, string theory has undergone many changes of direction, in the light of new evidence and discovery: 

1970s: Strong nuclear interactions  

1980s: Quantum gravity; ``theory of everything''  

1990s:  AdS/CFT: QCD (revival of 1970s); quark-gluon plasmas 

2000s:  AdS/CFT: superconductors  

2000s: Cosmic strings  

2010s: Fluid mechanics 

2010s:  Black hole/qubit correspondence: entanglement in Quantum  Information Theory   

For example, by stacking a large number of branes on top of one another, Juan Maldacena \cite{Maldacena} showed that a (D+1)-dimensional spacetime with all its gravitational interactions, may be dual to a non-gravitational theory that resides on its D-dimensional boundary. If this so-called holographic picture is correct, our universe maybe like Plato' s cave and we are the shadows projected on its walls. Its technical name is the ADS/CFT correspondence. Maldacena's 1998 ADS/CFT paper has garnered an incredible 7000+ citations. Interestingly enough, this is partly because it has found applications outside the traditional ``theory of everything'' milieu that one normally associates with string and M-theory. These, frequently serendipitous, applications include quark-gluon plasmas, high temperature superconductors and fluid mechanics. ADS/CFT is not the only branch of string/M-theory that has found applications in different areas of physics. After all, as shown above, string theory was originally invented in the 1970s to explain the behaviour of protons, neutrons and pions under the influence of the strong nuclear force.  

The partial nature of our understanding of string/M-theory has so far prevented any kind of smoking-gun experimental test in the fields of particle physics and cosmology. This has led some critics of string theory to suggest that it is not true science. This is easily refuted by studying the history of scientific discovery; the 30-year time lag between the EPR idea and Bell's falsifiable prediction provides a nice example. Nevertheless it cannot be denied that a prediction in string theory would be very welcome. Here we describe a prediction, not in the fields of particle physics and cosmology, but in quantum information theory.

\section{Four qubit entanglement: a falsifiable prediction}

Borsten, Dahanayake, Rubens and I at Imperial College teamed up with Alessio Marrani at CERN. We invoked this black hole-qubit/correspondence to predict a new result in quantum information theory. Noting that the classification of stringy black holes puts them in 31 different families, we predicted that four qubits can be entangled in 31 different ways \cite{Marrani}. By the way, this particular aspect of the correspondence is not a guess or a conjecture but a consequence of the Kostant-Sekiguchi theorem:

\begin{gather*}
		\text{Extremal black holes classification of $STU$ model}\\
		\updownarrow   \\
		\text{31 real nilpotent orbits of {$SO(4,4)$} acting on the $\rep{28}$} \\
		 \updownarrow  \\
		 \text{Kostant-Sekiguchi Correspondence} \\
		  \updownarrow  \\
		  \text{31 complex nilpotent orbits of {$SL(2)^{4}$} acting on the $\rep{(2,2,2,2)}$}\\
		   \updownarrow \\
		   \text{ 4 qubits entanglement classification }
\end{gather*}
This can, in principle, be tested in the laboratory and we are urging our experimental colleagues to find ways of doing just that. So the esoteric mathematics of string and M-theory might yet find practical applications. 
\section*{Acknowledgements}

I am grateful to my collaborators Leron Borsten, Duminda Dahanayake, Sergio Ferrara, Hajar Ibrahim, Alessio Marrani and William Rubens for their part in this research and especially to Leron Borsten for help with the manuscript.

{}
\end{document}